\documentclass{article}
\usepackage{amssymb}
\usepackage{graphicx}
\usepackage{amsmath}
\usepackage{a4wide}

\newtheorem{theorem}{Theorem}

\newtheorem{corollary}{Corollary}

\newtheorem{definition}{Definition}

\newtheorem{proposition}{Proposition}

\newenvironment{proof}[1][Proof]{\textbf{#1.} }{\ \rule{0.5em}{0.5em}}
\makeatletter
\def\@removefromreset#1#2{\let\@tempb\@elt
     \def\@tempa#1{@&#1}\expandafter\let\csname @*#1*\endcsname\@tempa
     \def\@elt##1{\expandafter\ifx\csname @*##1*\endcsname\@tempa\else
    \noexpand\@elt{##1}\fi}     \expandafter\edef\csname cl@#2\endcsname{\csname cl@#2\endcsname}     \let\@elt\@tempb
     \expandafter\let\csname @*#1*\endcsname\@undefined}

\@removefromreset{equation}{section}

\@removefromreset{theorem}{section}
\makeatother
\input{tcilatex}

\begin{document}

\title{Quantum states satisfying classical probability constraints}
\author{Elena R. Loubenets \\
Applied Mathematics Department, \\
Moscow State Institute of Electronics and Mathematics}
\maketitle

\begin{abstract}
For linear combinations of quantum product averages in an arbitrary
bipartite state, we derive new quantum Bell-form and CHSH-form inequalities
with the right-hand sides expressed in terms of a bipartite state. This
allows us to specify in a general setting bipartite state properties
sufficient for the validity of a classical CHSH-form inequality and the
perfect correlation form of the original Bell inequality for \emph{any}
bounded quantum observables. We also introduce a new general condition on a
bipartite state and quantum observables sufficient for the validity of the
original Bell inequality, in its perfect correlation or anticorrelation
forms. Under this general sufficient condition, a bipartite quantum state
does not necessarily exhibit perfect correlations or anticorrelations.
\end{abstract}

\tableofcontents

\section{Introduction}

The Bell [1] and the Clauser-Horne-Shimony-Holt (\emph{CHSH}) [2]
inequalities, derived originally in the frame of the Bell local hidden
variable model, describe the relations between the product expectation
values under different joint measurements.

In the frame of classical probability, for any bounded classical
observables, the product expectation values in every classical state satisfy
the original CHSH inequality and the perfect correlation form of the
original Bell inequality\footnote{%
The original proof [1] of the perfect correlation form of the Bell
inequality is true only for dichotomic classical observables admitting
values $\pm \lambda .$ In appendix of [3], we proved the validity of this
inequality for any three bounded classical observables.}.

In the frame of quantum probability and, more generally, quantum measurement
theory, the product expectation values under joint quantum measurements on a
bipartite system, do not, in general, satisfy a Bell-type inequality. It is,
however, well known [4,5] that there exist nonseparable bipartite states
that satisfy the CHSH inequality for any bounded quantum observables. We
also proved in [3] that there exist separable states\footnote{%
See [3], section 3.B.1.} that satisfy the perfect correlation form of the
original Bell inequality for any bounded quantum observables and do not
necessarily exhibit perfect correlations\footnote{%
The assumption of perfect correlations or anticorrelations, introduced by J.
Bell [1], represents a sufficient condition for the validity of the
corresponding form of the original Bell inequality for a bipartite quantum
state admitting a local hidden variable model.}.

At present, Bell-type inequalities are widely used in quantum information
processing. However, from the pioneering paper of R. Werner [4] up to now a
general analytical structure of bipartite quantum states satisfying a 
\textit{classical\footnote{%
In our paper, the term \textit{classical} specifies the validity of some
probabilistic constraint in the frame of classical probability.} }CHSH-form%
\textit{\ }inequality has not been well formalized. Moreover, a structure of
bipartite states satisfying the perfect correlation form of the original
Bell inequality for any bounded quantum observables has been analyzed (see
[3]) in the literature only in the separable case.

The aim of this paper is to introduce \emph{general analytical} \emph{%
conditions} sufficient for a bipartite quantum state to satisfy a \textit{%
classical }CHSH-form\textit{\ }inequality and the perfect correlation form
of the original Bell inequality for any quantum observables.

In\textit{\ section 2.1}, we introduce a new notion - a \emph{%
source-operator }for a bipartite quantum state and prove (proposition 1)
that, for any bipartite state, source-operators exist. We specify the new
notions - \emph{density source-operator (DSO) states}\footnote{%
In particular cases, specified in section 2.1, the notion of a density
source-operator reduces to the notion of a symmetric extension introduced in
[6].} and \emph{Bell class states, }and\emph{\ }present examples of such
bipartite states. We prove (proposition 2) that the nonseparable Werner state%
\footnote{%
This state was introduced by R.Werner in [4] and is widely used in quantum
information processing. See also the results on Werner states in [6] based
on the use of semi-definite programs.} is a DSO state for any dimension $%
d\geq 2$ and represents a Bell class state if $d\geq 3.$

In\textit{\ section 2.2}, we derive the \emph{new upper bounds}
(propositions 3, 4) of linear combinations of quantum product averages in an
arbitrary bipartite state. These quantum bounds are expressed in terms of
source-operators for a bipartite state, and this allows us to specify
analytically in \textit{section 3} the situations where a bipartite quantum
state satisfies a classical Bell-type inequality.

In \textit{section 3}:

(i) we prove (theorems 1, 2) that the product expectation values in a
density source-operator (DSO) state satisfy a \textit{classical} CHSH-form
inequality for any bounded quantum observables\footnote{%
Everywhere in this paper, quantum observables may have any spectral types.};

(ii) we prove (theorem 3, corollary 2) that every \emph{Bell class} \emph{%
state} satisfies a \textit{classical} CHSH-form inequality and the perfect
correlation form of the original Bell inequality for any bounded quantum
observables and does not necessarily exhibit\ perfect correlations;

(iii)\ we introduce a new \emph{general condition }(theorem 4)\emph{\ }%
sufficient\emph{\ }for a density source-operator (DSO) state and three
bounded quantum observables to satisfy the original Bell inequality, in its
perfect correlation or anticorrelation forms. A DSO state, satisfying this
general sufficient condition does not necessarily exhibit (proposition 5)
Bell's perfect correlations/anticorrelations [1].

In\textit{\ section 4}, we specify (theorems 5 - 7) the validity of
classical Bell-type inequalities under generalized quantum measurements of
Alice and Bob.

\section{Quantum upper bounds. General case}

Let a bipartite quantum system be described in terms of a separable complex
Hilbert space $\mathcal{H}_{1}\otimes \mathcal{H}_{2}$. In this section, for
an arbitrary state\footnote{%
We consider only normal quantum states.} $\rho $ on $\mathcal{H}_{1}\otimes 
\mathcal{H}_{2},$ we derive the new upper bounds of linear combinations of
quantum product averages: 
\begin{eqnarray}
&&\mathrm{tr}[\rho (W_{1}^{(a)}\otimes W_{2}^{(b_{1})})]-\mathrm{tr}[\rho
(W_{1}^{(a)}\otimes W_{2}^{(b_{2})})],\text{ }  \label{1} \\
&&\mathrm{tr}[\rho (W_{1}^{(a_{1})}\otimes W_{2}^{(b)})]-\mathrm{tr}[\rho
(W_{1}^{(a_{2})}\otimes W_{2}^{(b)})],  \notag \\
&&\sum_{n,m}\gamma _{nm}\mathrm{tr}[\rho (W_{1}^{(a_{n})}\otimes
W_{2}^{(b_{m})})],  \label{2}
\end{eqnarray}%
Here, $W_{1}^{(a)},$ $W_{2}^{(b)}$ are any bounded quantum observables on $%
\mathcal{H}_{1}$ and $\mathcal{H}_{2},$ respectively, and $\gamma _{nm}$, $%
n,m=1,2,$ are any real coefficients. For clearness, we label\footnote{%
In the physical literature, these labels correspond to "Alice" and "Bob"
names.} by indices "$a$" quantum observables on $\mathcal{H}_{1}$ and by $%
"b" $- on $\mathcal{H}_{2}.$

\subsection{Source-operators for a bipartite state}

In order to evaluate (\ref{1}) and (\ref{2}), we introduce in a general
setting a new notion.

Denote by $\mathcal{K}_{112}:=\mathcal{H}_{1}\otimes \mathcal{H}_{1}\otimes 
\mathcal{H}_{2}$ and $\mathcal{K}_{122}:=\mathcal{H}_{1}\otimes \mathcal{H}%
_{2}\otimes \mathcal{H}_{2}$ the extended tensor product Hilbert spaces.
Below, we use the notation $\mathrm{tr}_{\mathcal{H}_{m}}^{(k)}[\cdot ],$ $%
k=1,2,3;$ $m=1,2$, for the partial trace over the elements of a Hilbert
space $\mathcal{H}_{m}$ standing in the $k$-th place of tensor products.

\begin{definition}[Source-operators]
For a state $\rho $ on $\mathcal{H}_{1}\otimes \mathcal{H}_{2}$, let $%
T_{112} $ on $\mathcal{K}_{112}$ and $T_{122}$ on $\mathcal{K}_{122}$ be
self-adjoint trace class operators defined by the relations: 
\begin{eqnarray}
\mathrm{tr}_{\mathcal{H}_{1}}^{(1)}[T_{112}] &=&\rho ,\text{ \ \ \ }\mathrm{%
tr}_{\mathcal{H}_{1}}^{(2)}[T_{112}]=\rho ;  \label{3} \\
\mathrm{tr}_{\mathcal{H}_{2}}^{(2)}[T_{122}] &=&\rho ,\text{ \ \ \ }\mathrm{%
tr}_{\mathcal{H}_{2}}^{(3)}[T_{122}]=\rho .  \label{4}
\end{eqnarray}
We call any of these dilations a source-operator\textit{\ for} a bipartite
state $\rho $.
\end{definition}

\begin{proposition}
For a state $\rho $ on $\mathcal{H}_{1}\mathcal{\otimes H}_{2}$, there exist
source-operators $T_{122}$ and $T_{112}$.
\end{proposition}

\begin{proof}
The spectral decomposition of a quantum state $\rho $ on $\mathcal{H}%
_{1}\otimes \mathcal{H}_{2}$ reads: 
\begin{equation}
\rho =\sum_{i}\alpha _{i}|\Psi _{i}\rangle \langle \Psi _{i}|,\text{ \ \ }%
\langle \Psi _{i},\Psi _{j}\rangle =\delta _{ij}\text{, \ \ \ }\forall
\alpha _{i}>0,\text{ \ \ }\sum_{i}\alpha _{i}=1.  \label{5}
\end{equation}%
Take an orthonormal basis $\{\varphi _{n}\}$ in $\mathcal{H}_{2}$ and
consider the Schmidt decomposition of an eigenvector\ $\Psi _{i}$ with
respect to this basis: 
\begin{equation}
\Psi _{i}=\sum_{n}\Phi _{n}^{(i)}\otimes \varphi _{n},\ \text{\ \ \ }%
\sum_{n}\langle \Phi _{n}^{(i)},\Phi _{n}^{(j)}\rangle =\delta _{ij}.
\label{6}
\end{equation}%
Substituting (\ref{6}) into (\ref{5}), we derive $\rho =\sum_{n,m}\rho
_{nm}\otimes |\varphi _{n}\rangle \langle \varphi _{m}|,$ where $\rho
_{nm}:=\sum_{i}\alpha _{i}|\Phi _{n}^{(i)}\rangle \langle \Phi _{m}^{(i)}|,$ 
$\forall n,m.$ The operators $\rho _{nn}$ are positive with $\sum_{n}\mathrm{%
tr}[\rho _{nn}]=1.$\newline
For any density operator $\sigma $ on $\mathcal{H}_{2}$ and any self-adjoint
trace class operator $\tau _{122}$ on $\mathcal{K}_{122}$, with $\mathrm{tr}%
_{\mathcal{H}_{2}}^{(2)}[\tau _{122}]=\mathrm{tr}_{\mathcal{H}%
_{2}}^{(3)}[\tau _{122}]=0,$ the operator 
\begin{eqnarray}
T_{122} &=&\sum_{n,m}\rho _{nm}\otimes |\varphi _{n}\rangle \langle \varphi
_{m}|\otimes \sigma +\sum_{n,m}\rho _{nm}\otimes \sigma \otimes |\varphi
_{n}\rangle \langle \varphi _{m}|\text{ }  \label{7} \\
&&-\text{ }\mathrm{tr}_{\mathcal{H}_{2}}[\rho ]\otimes \sigma \otimes \sigma 
\text{ }+\tau _{122}  \notag
\end{eqnarray}%
represents a source-operator for the state (\ref{5}). Here, \textrm{tr}$_{%
\mathcal{H}_{2}}[\rho ]=\sum_{n}\rho _{nn}$ is the density operator on $%
\mathcal{H}_{1}$ reduced from $\rho .$ The existence of a source-operator $%
T_{112}$ is proved similarly.
\end{proof}

\smallskip

\noindent Consider now the main properties of source-operators:

\begin{enumerate}
\item $\mathrm{tr}[T]=1,$ for any source-operator $T;$

\item As any self-adjoint trace class operator, a source-operator admits the
decomposition $T=T^{(+)}-T^{(-)}$ via non-negative operators $T^{(+)}=\frac{1%
}{2}(|T|+T)$ and $T^{(-)}=\frac{1}{2}(|T|-T),$ and $||T||_{1}=\mathrm{tr}%
[T^{(+)}]+\mathrm{tr}[T^{(-)}].$ For a source-operator, the latter relation
implies $||T||_{1}=1+2\mathrm{tr}[T^{(-)}];$

\item Any positive source-operator $T$ is a density operator and we refer to
it as a density source-operator\emph{\ (DSO). }A source-operator\emph{\ }is%
\emph{\ }a DSO iff $\left\Vert T\right\Vert _{1}=1.$
\end{enumerate}

\smallskip

\begin{definition}[DSO states]
\footnote{%
If, in particular, a density source-operator, for example, $T_{122}$ is
symmetric with respect to the permutation of elements standing in the second
and the third places of tensor products on $\mathcal{K}_{122},$ then this $%
T_{122}$ represents a (1,2) symmetric extension in the terminology
introduced in [6].}If a bipartite state has a density source-operator then
we call this state as a density source-operator state or a \textit{DSO}
state, for short.
\end{definition}

Consider a separable state $\rho _{sep}.$ Let $\sum_{m}\xi _{m}\rho
_{1}^{(m)}\otimes \rho _{2}^{(m)}$, where $\xi _{m}>0,$ $\sum_{m}\xi _{m}=1,$
be a separable representation of $\rho _{sep}.$ Then, for example, $%
T_{122}=\sum_{m}\xi _{m}\rho _{1}^{(m)}\otimes \rho _{2}^{(m)}\otimes \rho
_{2}^{(m)}$\ is a density source-operator for $\rho _{sep}.$

Hence, \emph{any separable state is a DSO state}. However, the converse is
not true and a DSO state may be nonseparable. In section 2.1.1, we consider
examples of nonseparable DSO states, in particular, on infinite dimensional
Hilbert space.

If $\mathcal{H}_{1}=\mathcal{H}_{2}=\mathcal{H}$ then $\mathcal{K}_{122}=%
\mathcal{K}_{112}=\mathcal{H}\otimes \mathcal{H}\otimes \mathcal{H}$ and in
order to distinguish between source-operators $T_{112}$ and $T_{122}$ we
further label\footnote{%
These labels indicate a \textquotedblright direction\textquotedblright\ of
dilation.} them as $T_{\blacktriangleleft }$ and $T_{\blacktriangleright },$
respectively. Moreover, if there exists a source-operator that satisfies
both conditions in definition 1, then we denote this it by $%
T_{\blacktriangleleft \blacktriangleright }$. The latter source-operator%
\emph{\ }has the special dilation property: 
\begin{equation}
\mathrm{tr}_{\mathcal{H}}^{(1)}[T_{\blacktriangleleft \blacktriangleright }]=%
\mathrm{tr}_{\mathcal{H}}^{(2)}[T_{\blacktriangleleft \blacktriangleright }]=%
\mathrm{tr}_{\mathcal{H}}^{(3)}[T_{\blacktriangleleft \blacktriangleright
}]=\rho .  \label{8}
\end{equation}

\begin{definition}[Bell class states]
If, for a density source-operator (DSO) state on $\mathcal{H}\otimes 
\mathcal{H}$ , there exists a density source-operator with the special
dilation property (\ref{8}) then we refer to this DSO state as a Bell class
state.
\end{definition}

The Bell class includes both separable and nonseparable states. Separable
states on $\mathcal{H}\otimes \mathcal{H}$ of the special form (49)
introduced in [3], namely: $\sum_{m}\xi _{m}\rho ^{(m)}\otimes \rho ^{(m)},$ 
$\xi _{m}>0,$ $\sum_{m}\xi _{m}=1,$ constitute examples of separable Bell
class states.

\subsubsection{Examples of DSO and Bell class states}

In this section, we present examples of nonseparable DSO and Bell class
states on $\mathcal{H}\otimes \mathcal{H}.$

Consider the nonseparable Werner state [4] 
\begin{equation}
\rho _{W}^{(d)}=\frac{d+1}{d^{3}}I_{\mathbb{C}^{d}\otimes \mathbb{C}^{d}}-%
\frac{1}{d^{2}}V_{d}  \label{9}
\end{equation}
on $\mathbb{C}^{d}\otimes \mathbb{C}^{d},$ $d\geq 2.$ Here, $V_{d}$ is the
permutation operator: $V_{d}(\psi _{1}\otimes \psi _{2}):=\psi _{2}\otimes
\psi _{1},$ $\forall \psi _{1},\psi _{2}\in \mathbb{C}^{d}.$ This operator
is self-adjoint and has the properties: $(V_{d})^{2}=I_{\mathbb{C}%
^{d}\otimes \mathbb{C}^{d}}$, $\mathrm{tr}[V_{d}]=d.$

\begin{proposition}
The nonseparable Werner state $\rho _{W}^{(d)},$ $\forall d\geq 2,$
represents a DSO state and is of the Bell class for any $d\geq 3.$
\end{proposition}

\begin{proof}
Introduce on $\mathbb{C}^{d}\otimes \mathbb{C}^{d},$ $\forall d\geq 3,$ the
orthogonal projection 
\begin{eqnarray}
Q_{d}^{(-)}(\psi _{1}\otimes \psi _{2}\otimes \psi _{3}) &:&=\frac{1}{6}%
\{\psi _{1}\otimes \psi _{2}\otimes \psi _{3}-\psi _{2}\otimes \psi
_{1}\otimes \psi _{3}-\psi _{1}\otimes \psi _{3}\otimes \psi _{2}  \label{10}
\\
&&-\psi _{3}\otimes \psi _{2}\otimes \psi _{1}+\psi _{2}\otimes \psi
_{3}\otimes \psi _{1}+\psi _{3}\otimes \psi _{1}\otimes \psi _{2}\},  \notag
\end{eqnarray}%
$\forall \psi _{1},\psi _{2},\psi _{3}\in \mathbb{C}^{d}.$ This projection
has the form: 
\begin{eqnarray}
6Q_{d}^{(-)} &=&I_{\mathbb{C}^{d}\otimes \mathbb{C}^{d}\otimes \mathbb{C}%
^{d}}-V_{d}\otimes I_{\mathbb{C}^{d}}-I_{\mathbb{C}^{d}}\otimes V_{d}-(I_{%
\mathbb{C}^{d}}\otimes V_{d})(V_{d}\otimes I_{\mathbb{C}^{d}})(I_{\mathbb{C}%
^{d}}\otimes V_{d})  \label{11} \\
&&+(I_{\mathbb{C}^{d}}\otimes V_{d})(V_{d}\otimes I_{\mathbb{C}%
^{d}})+(V_{d}\otimes I_{\mathbb{C}^{d}})(I_{\mathbb{C}^{d}}\otimes V_{d}) 
\notag
\end{eqnarray}%
and admits a representation: 
\begin{eqnarray}
6Q_{d}^{(-)} &=&I_{\mathbb{C}^{d}\otimes \mathbb{C}^{d}\otimes \mathbb{C}%
^{d}}-\tsum_{n,m}|e_{n}\rangle \langle e_{m}|\otimes |e_{m}\rangle \langle
e_{n}|\otimes I_{\mathbb{C}^{d}}-\tsum_{n,m}I_{\mathbb{C}^{d}}\otimes
|e_{n}\rangle \langle e_{m}|\otimes |e_{m}\rangle \langle e_{n}|  \label{12}
\\
&&-\tsum_{n,m}|e_{n}\rangle \langle e_{m}|\otimes I_{\mathbb{C}^{d}}\otimes
|e_{m}\rangle \langle e_{n}|\text{ }+\tsum_{n,m,k}|e_{n}\rangle \langle
e_{m}|\otimes |e_{m}\rangle \langle e_{k}|\otimes |e_{k}\rangle \langle
e_{n}|\text{ }  \notag \\
&&+\tsum_{n,m,k}|e_{m}\rangle \langle e_{n}|\otimes |e_{k}\rangle \langle
e_{m}|\otimes |e_{n}\rangle \langle e_{k}|  \notag
\end{eqnarray}%
via an orthonormal basis $\{e_{n}\}$ in $\mathbb{C}^{d}$ (notice that $%
V_{d}=\tsum_{n,m=1}^{d}|e_{n}\rangle \langle e_{m}|\otimes |e_{m}\rangle
\langle e_{n}|$ ).\newline
We have $\mathrm{tr}_{\mathbb{C}^{d}}^{(j)}[Q_{d}^{(-)}]=\frac{d-2}{6}(I_{%
\mathbb{C}^{d}\otimes \mathbb{C}^{d}}-V_{d}),$ $\forall j=1,2,3.$ Hence, for
the state $\rho _{W}^{(d)},$ $\forall d\geq 3,$ the operator 
\begin{equation}
R_{\blacktriangleleft \blacktriangleright }^{(d)}=\frac{1}{d^{4}}I_{\mathbb{C%
}^{d}\otimes \mathbb{C}^{d}\otimes \mathbb{C}^{d}}+\frac{6}{d^{2}(d-2)}%
Q_{d}^{(-)}  \label{13}
\end{equation}%
represents a density source-operator with the special dilation property (8),
that is: $\mathrm{tr}_{\mathbb{C}^{d}}^{(j)}[R_{\blacktriangleleft
\blacktriangleright }^{(d)}]=\rho _{W}^{(d)},$ $\forall j=1,2,3.$ If $d=2,$
then 
\begin{equation}
R_{\blacktriangleright }^{(2)}=\frac{1}{4}I_{\mathbb{C}^{2}\otimes \mathbb{C}%
^{2}\otimes \mathbb{C}^{2}}-\frac{1}{8}V_{2}\otimes I_{\mathbb{C}^{2}}-\frac{%
1}{8}(I_{\mathbb{C}^{2}}\otimes V_{2})(V_{2}\otimes I_{\mathbb{C}^{2}})(I_{%
\mathbb{C}^{2}}\otimes V_{2})  \label{14}
\end{equation}%
is a density source-operator for $\rho _{W}^{(2)}.$ The existence of the
density source-operators (\ref{13}) and (\ref{14}) proves the statement.
\end{proof}

\smallskip

Consider now examples of DSO and Bell class states on infinite dimensional
Hilbert space $\mathcal{H}\otimes \mathcal{H}$. Take the quantum states 
\begin{eqnarray}
\rho _{1} &=&\frac{1}{4}|\psi _{1}\otimes \psi _{1}+\psi _{2}\otimes \psi
_{2}\rangle \langle \psi _{1}\otimes \psi _{1}+\psi _{2}\otimes \psi _{2}|
\label{15} \\
&&+\frac{1}{4}(|\psi _{1}\rangle \langle \psi _{1}|+|\psi _{2}\rangle
\langle \psi _{2}|)\otimes |\psi _{1}\rangle \langle \psi _{1}|  \notag
\end{eqnarray}%
and 
\begin{eqnarray}
\rho _{2} &=&\frac{1}{6}|\psi _{1}\otimes \psi _{1}+\psi _{2}\otimes \psi
_{2}\rangle \langle \psi _{1}\otimes \psi _{1}+\psi _{2}\otimes \psi _{2}|
\label{16} \\
&&+\frac{1}{6}(|\psi _{1}\rangle \langle \psi _{1}|+|\psi _{2}\rangle
\langle \psi _{2}|)\otimes |\psi _{1}\rangle \langle \psi _{1}|  \notag \\
&&+\frac{1}{6}|\psi _{1}\rangle \langle \psi _{1}|\otimes (|\psi _{1}\rangle
\langle \psi _{1}|+|\psi _{2}\rangle \langle \psi _{2}|),  \notag
\end{eqnarray}%
where $\psi _{1},\psi _{2}\in \mathcal{H}$ are any mutually orthogonal unit
vectors. The partial transpose $\rho _{1}^{T_{1}}$\ has the negative
eigenvalue $\lambda =\frac{1}{8}(1-\sqrt{5})$, corresponding to the
eigenvector $\psi =c(\psi _{1}\otimes \psi _{2}+\frac{1-\sqrt{5}}{2}\psi
_{2}\otimes \psi _{1}).$ Therefore, due to the Peres separability criterion
[7], the state $\rho _{1}$ is nonseparable. Nonseparability of $\rho _{2}$
is proved similarly. The operators 
\begin{eqnarray}
R_{\blacktriangleright } &=&\frac{1}{4}|\psi _{1}\otimes \psi _{1}+\psi
_{2}\otimes \psi _{2}\rangle \langle \psi _{1}\otimes \psi _{1}+\psi
_{2}\otimes \psi _{2}|\otimes |\psi _{1}\rangle \langle \psi _{1}|
\label{17} \\
&&+\frac{1}{4}|\text{ }\psi _{1}\otimes \psi _{1}\otimes \psi _{1}+\psi
_{2}\otimes \psi _{1}\otimes \psi _{2}\rangle \langle \psi _{1}\otimes \psi
_{1}\otimes \psi _{1}+\psi _{2}\otimes \psi _{1}\otimes \psi _{2}|  \notag
\end{eqnarray}%
and 
\begin{eqnarray}
R_{\blacktriangleleft \blacktriangleright } &=&\frac{1}{6}|\psi _{1}\otimes
\psi _{1}+\psi _{2}\otimes \psi _{2}\rangle \langle \psi _{1}\otimes \psi
_{1}+\psi _{2}\otimes \psi _{2}|\otimes |\psi _{1}\rangle \langle \psi _{1}|
\label{18} \\
&&+\frac{1}{6}|\text{ }\psi _{1}\otimes \psi _{1}\otimes \psi _{1}+\psi
_{2}\otimes \psi _{1}\otimes \psi _{2}\rangle \langle \psi _{1}\otimes \psi
_{1}\otimes \psi _{1}+\psi _{2}\otimes \psi _{1}\otimes \psi _{2}|  \notag \\
&&+\frac{1}{6}|\text{ }\psi _{1}\otimes \psi _{1}\otimes \psi _{1}+\psi
_{1}\otimes \psi _{2}\otimes \psi _{2}\rangle \langle \psi _{1}\otimes \psi
_{1}\otimes \psi _{1}+\psi _{1}\otimes \psi _{2}\otimes \psi _{2}|  \notag
\end{eqnarray}%
represent density source-operators for $\rho _{1}$ and $\rho _{2},$
respectively. Moreover, the DSO $R_{\blacktriangleleft \blacktriangleright }$
has the special dilation property (\ref{8}). Hence: (i) $\rho _{1}$ is a
nonseparable DSO state; (ii) $\rho _{2}$ is a nonseparable Bell class state.

\subsection{Quantum Bell-form inequalities}

Based on the new notion of a source-operator, consider now upper bounds of
linear combinations (\ref{1}) of quantum product averages in an arbitrary
state $\rho $ on $\mathcal{H}_{1}\otimes \mathcal{H}_{2}$.

Let $T_{122}$ and $T_{112}$ be any source-operators for a state $\rho $.
According to proposition 1, for any bipartite state $\rho $, these operators
exist.

In view of definition 1, we have: 
\begin{eqnarray}
&&\mathrm{tr}[\rho (W_{1}^{(a)}\otimes W_{2}^{(b_{1})}-W_{1}^{(a)}\otimes
W_{2}^{(b_{2})})]  \label{19} \\
&=&\mathrm{tr}[T_{122}(W_{1}^{(a)}\otimes W_{2}^{(b_{1})}\otimes I_{\mathcal{%
H}_{2}}-W_{1}^{(a)}\otimes I_{\mathcal{H}_{2}}\otimes W_{2}^{(b_{2})})], 
\notag \\
&&  \notag \\
&&\mathrm{tr}[\rho (W_{1}^{(a_{1})}\otimes
W_{2}^{(b)}-W_{1}^{(a_{2})}\otimes W_{2}^{(b)})]  \notag \\
&=&\mathrm{tr}[T_{112}(W_{1}^{(a_{1})}\otimes I_{\mathcal{H}_{1}}\otimes
W_{2}^{(b)}-I_{\mathcal{H}_{1}}\otimes W_{1}^{(a_{2})}\otimes W_{2}^{(b)})],
\notag
\end{eqnarray}%
and these representations allow us to prove the following general statement.

\begin{proposition}
Let $W_{1}^{(a_{n})},$ $W_{2}^{(b_{n})},$ $n=1,2$ be any bounded quantum
observables with operator norms $||\cdot ||\leq 1.$ An arbitrary state $\rho 
$ on $\mathcal{H}_{1}\otimes \mathcal{H}_{2}$ satisfies the inequalities 
\begin{eqnarray}
\left| \mathrm{tr}[\rho (W_{1}^{(a)}\otimes W_{2}^{(b_{1})})]-\mathrm{tr}%
[\rho (W_{1}^{(a)}\otimes W_{2}^{(b_{2})})]\right|  \label{20} \\
\leq ||T_{122}||_{1}\text{ }\{1-\mathrm{tr}[\sigma
_{T_{122}}(W_{2}^{(b_{1})}\otimes W_{2}^{(b_{2})})]\text{ }\}  \notag
\end{eqnarray}
and 
\begin{eqnarray}
\left| \mathrm{tr}[\rho (W_{1}^{(a_{1})}\otimes W_{2}^{(b)})]-\mathrm{tr}%
[\rho (W_{1}^{(a_{2})}\otimes W_{2}^{(b)})]\right|  \label{21} \\
\leq ||T_{112}||_{1}\text{ }\{1-\mathrm{tr}[\sigma
_{T_{112}}(W_{1}^{(a_{1})}\otimes W_{1}^{(a_{2})})]\text{ }\},  \notag
\end{eqnarray}
where $T_{122}$ and $T_{112}$ are any source-operators for $\rho $ and 
\begin{equation}
\sigma _{T_{122}}:=\frac{1}{||T_{122}||_{1}}\mathrm{tr}_{\mathcal{H}%
_{1}}^{(1)}[|T_{122}|]\text{, \ \ \ \ }\sigma _{T_{112}}:=\frac{1}{%
||T_{112}||_{1}}\mathrm{tr}_{\mathcal{H}_{2}}^{(3)}[|T_{112}|]  \label{22}
\end{equation}
are density operators on $\mathcal{H}_{2}\otimes \mathcal{H}_{2}$ and $%
\mathcal{H}_{1}\otimes \mathcal{H}_{1},$ respectively. \newline
In the right-hand side of (\ref{20}) (or (\ref{21})), the observables can be
interchanged.
\end{proposition}

\begin{proof}
In order to prove (\ref{20}), we notice that in (\ref{19}): 
\begin{eqnarray}
W_{1}^{(a)}\otimes W_{2}^{(b_{1})}\otimes I_{\mathcal{H}_{2}}
&=&(W_{1}^{(a)}\otimes I_{\mathcal{H}_{2}}\otimes I_{\mathcal{H}_{2}})(I_{%
\mathcal{H}_{1}}\otimes W_{2}^{(b_{1})}\otimes I_{\mathcal{H}_{2}}),
\label{23} \\
W_{1}^{(a)}\otimes I_{\mathcal{H}_{2}}\otimes W_{2}^{(b_{2})}
&=&(W_{1}^{(a)}\otimes I_{\mathcal{H}_{2}}\otimes I_{\mathcal{H}_{2}})(\text{%
\ }I_{\mathcal{H}_{1}}\otimes I_{\mathcal{H}_{2}}\otimes W_{2}^{(b_{2})}), 
\notag
\end{eqnarray}%
and the bounded quantum observables 
\begin{equation}
W_{1}^{(a)}\otimes I_{\mathcal{H}_{2}}\otimes I_{\mathcal{H}_{2}},\text{ \ \
\ }I_{\mathcal{H}_{1}}\otimes W_{2}^{(b_{1})}\otimes I_{\mathcal{H}_{2}},%
\text{ \ \ \ }I_{\mathcal{H}_{1}}\otimes I_{\mathcal{H}_{2}}\otimes
W_{2}^{(b_{2})}  \label{24}
\end{equation}%
on $\mathcal{K}_{122}$ mutually commute. From the von Neumann theorem ([8],
page 221) it follows that \emph{there exist}: \newline
(i) a bounded quantum observable $V_{a}^{(b_{1},b_{2})}$ on $\mathcal{K}%
_{122}$; \newline
(ii) bounded Borel real-valued functions $\varphi _{1}^{(a)},\varphi
_{2}^{(b_{1})},\varphi _{3}^{(b_{2})}$ on $(\mathbb{R},$ $\mathcal{B}_{%
\mathbb{R}}),$ with supremum norms $||\varphi _{1}^{(a)}||,$ $||\varphi
_{2}^{(b_{1})}||,$ $||\varphi _{3}^{(b_{2})}||$ $\leq 1;$\newline
such that 
\begin{eqnarray}
W_{1}^{(a)}\otimes I_{\mathcal{H}_{2}}\otimes I_{\mathcal{H}_{2}} &=&\varphi
_{1}^{(a)}(V_{a}^{(b_{1},b_{2})}),\text{ \ \ }I_{\mathcal{H}_{1}}\otimes
W_{2}^{(b_{1})}\otimes I_{\mathcal{H}_{2}}=\varphi
_{2}^{(b_{1})}(V_{a}^{(b_{1},b_{2})}),  \label{25} \\
I_{\mathcal{H}_{1}}\otimes I_{\mathcal{H}_{2}}\otimes W_{2}^{(b_{2})}
&=&\varphi _{3}^{(b_{2})}(V_{a}^{(b_{1},b_{2})}).  \notag
\end{eqnarray}%
Let $P_{V_{a}^{(b_{1},b_{2})}}(\cdot ),$ $P_{V_{a}^{(b_{1},b_{2})}}(\mathbb{R%
})=I_{\mathcal{K}_{122}},$ be the projection-valued measure corresponding
uniquely to $V_{a}^{(b_{1},b_{2})}$ due to the spectral theorem. In view of (%
\ref{19}) and (\ref{25}), 
\begin{eqnarray}
\mathrm{tr}[\rho (W_{1}^{(a)}\otimes W_{2}^{(b_{1})})] &=&\int_{\mathbb{R}%
}\varphi _{1}^{(a)}(\xi )\varphi _{2}^{(b_{1})}(\xi )\nu
_{a}^{(b_{1},b_{2})}(d\xi ;T_{122}),  \label{26} \\
\mathrm{tr}[\rho (W_{1}^{(a)}\otimes W_{2}^{(b_{2})})] &=&\int_{\mathbb{R}%
}\varphi _{1}^{(a)}(\xi )\varphi _{3}^{(b_{2})}(\xi )\nu
_{a}^{(b_{1},b_{2})}(d\xi ;T_{122}),  \notag
\end{eqnarray}%
where we denote by $\nu _{a}^{(b_{1},b_{2})}(\cdot ;Y)$ a $\sigma $-additive
bounded real-valued measure on $(\mathbb{R},$ $\mathcal{B}_{\mathbb{R}}),$
defined by the relation 
\begin{equation}
\nu _{a}^{(b_{1},b_{2})}(\cdot ;Y):=\mathrm{tr}[YP_{V_{a}^{(b_{1},b_{2})}}(%
\cdot )],\text{ \ \ }\nu _{a}^{(b_{1},b_{2})}(\mathbb{R};Y)=\mathrm{tr}[Y],
\label{27}
\end{equation}%
for any self-adjoint trace class operator $Y$ on $\mathcal{K}_{122}.$ For a
source-operator $T_{122},$ the measure $\nu _{a}^{(b_{1},b_{2})}(\cdot
;T_{122})$ is normalized but not, in general, positive. Due to property 2,
section 2.1, 
\begin{equation}
\nu _{a}^{(b_{1},b_{2})}(\cdot ;T_{122})=\nu _{a}^{(b_{1},b_{2})}(\cdot
;T_{122}^{(+)})-\nu _{a}^{(b_{1},b_{2})}(\cdot ;T_{122}^{(-)}),  \label{28}
\end{equation}%
where $\nu _{a}^{(b_{1},b_{2})}(\cdot ;T_{122}^{(\pm )})$ are unnormalized
positive measures with 
\begin{eqnarray}
&&\nu _{a}^{(b_{1},b_{2})}(\mathbb{R};T_{122}^{(+)})+\nu
_{a}^{(b_{1},b_{2})}(\mathbb{R};T_{122}^{(-)})  \label{29} \\
&=&\nu _{a}^{(b_{1},b_{2})}(\mathbb{R};|T_{122}|)  \notag \\
&=&||T_{122}||_{1}.  \notag
\end{eqnarray}%
Using (\ref{26}), (\ref{28}), the bound $|\varphi _{1}^{(a)}||\leq 1$, and
the inequality $|x-y|\leq 1-xy,$ valid for any real numbers $|x|\leq 1,$ $%
|y|\leq 1,$ we derive: 
\begin{eqnarray}
&&\left\vert \mathrm{tr}[\rho (W_{1}^{(a)}\otimes
W_{2}^{(b_{1})}-W_{1}^{(a)}\otimes W_{2}^{(b_{2})})]\right\vert  \label{30}
\\
&\leq &\nu _{a}^{(b_{1},b_{2})}(\mathbb{R};|T_{122}|)-\int_{\mathbb{R}%
}\varphi _{2}^{(b_{1})}(\xi )\varphi _{3}^{(b_{2})}(\xi )\nu
_{a}^{(b_{1},b_{2})}(d\mathbb{\xi };|T_{122}|).  \notag
\end{eqnarray}%
Due to (\ref{25}) and (\ref{27}), 
\begin{eqnarray}
&&\int_{\mathbb{R}}\varphi _{2}^{(b_{1})}(\xi )\varphi _{3}^{(b_{2})}(\xi
)\nu _{a}^{(b_{1},b_{2})}(d\mathbb{\xi };|T_{122}|)  \label{31} \\
&=&||T_{122}||_{1}\mathrm{tr}[\sigma _{T_{122}}(W_{2}^{(b_{1})}\otimes
W_{2}^{(b_{2})})]\text{,}  \notag
\end{eqnarray}%
where $\sigma _{T_{122}}:=\frac{1}{||T_{122}||_{1}}\mathrm{tr}_{\mathcal{H}%
_{1}}^{(1)}[$ $|T_{122}|$ $]$ is a density operator on $\mathcal{H}%
_{2}\otimes \mathcal{H}_{2}.$ \newline
Substituting (\ref{29}) and (\ref{31}) into (\ref{30}), we finally have: 
\begin{eqnarray}
&&\left\vert \mathrm{tr}[\rho (W_{1}^{(a)}\otimes
W_{2}^{(b_{1})}-W_{1}^{(a)}\otimes W_{2}^{(b_{2})})]\right\vert  \label{32}
\\
&\leq &\left\Vert T_{122}\right\Vert _{1}\{1-\mathrm{tr}[\sigma
_{T_{122}}(W_{2}^{(b_{1})}\otimes W_{2}^{(b_{2})})]\text{ }\}.  \notag
\end{eqnarray}%
The derivation of the inequality (\ref{21}) is quite similar.
\end{proof}

\begin{corollary}
Let $W_{1}^{(a)}$ and $W_{2}^{(b)}$ be any bounded quantum observables with
operator norms $||\cdot ||\leq 1.$ For any state $\rho $ on $\mathcal{H}%
_{1}\otimes \mathcal{H}_{2},$ the inequalities: 
\begin{eqnarray}
\left\vert \text{ }\mathrm{tr}[\rho (W_{1}^{(a)}\otimes
W_{2}^{(b)})]\right\vert &\leq &\frac{1}{2}||T_{122}||_{1}\text{ }\{1+%
\mathrm{tr}[\sigma _{T_{122}}(W_{2}^{(b)}\otimes W_{2}^{(b)})]\text{ }\},
\label{33} \\
\left\vert \text{ }\mathrm{tr}[\rho (W_{1}^{(a)}\otimes
W_{2}^{(b)})]\right\vert &\leq &\frac{1}{2}||T_{112}||_{1}\text{ }\{1+%
\mathrm{tr}[\sigma _{T_{112}}(W_{1}^{(a)}\otimes W_{1}^{(a)})]\text{ }\} 
\notag
\end{eqnarray}%
hold with arbitrary source-operators $T_{122}$ and $T_{112}$ for $\rho $ in
the right hand sides.
\end{corollary}

In particular, for a Bell class state $\rho $ on $\mathcal{H}\otimes 
\mathcal{H}$, the relations (\ref{33}) imply: 
\begin{eqnarray}
\left\vert \text{ }\mathrm{tr}[\rho (W_{1}\otimes W_{2})]\right\vert &\leq &%
\frac{1}{2}\{1+\mathrm{tr}[\rho (W_{2}\otimes W_{2})]\text{ }\},  \label{34}
\\
\left\vert \text{ }\mathrm{tr}[\rho (W_{1}\otimes W_{2})]\right\vert &\leq &%
\frac{1}{2}\{1+\mathrm{tr}[\rho (W_{1}\otimes W_{1})]\text{ }\},  \notag
\end{eqnarray}%
for any $W_{1}$ and $W_{2}$ on $\mathcal{H}.$

\subsection{Quantum CHSH-form inequalities}

Consider now upper bounds for a linear combination (\ref{2}).

\begin{proposition}
Let $W_{1}^{(a_{n})}$ and $W_{2}^{(b_{m})}$, $n,m=1,2,$ be any bounded
quantum observables with operator norms $||\cdot ||\leq 1$ and $\gamma _{nm}$%
, $n,m=1,2$, be any real coefficients with $\left\vert \gamma
_{nm}\right\vert \leq 1.$ \newline
An arbitrary quantum state $\rho $ on $\mathcal{H}_{1}\otimes \mathcal{H}%
_{2} $ satisfies the inequality 
\begin{equation}
\text{\ }\left\vert \sum_{n,m=1,2}\gamma _{nm}\mathrm{tr}[\rho
(W_{1}^{(a_{n})}\otimes W_{2}^{(b_{m})})]\right\vert \leq 2\left\Vert
T_{122}\right\Vert _{1},  \label{35}
\end{equation}%
whenever $\gamma _{11}\gamma _{12}=-\gamma _{21}\gamma _{22},$ and the
inequality\ 
\begin{equation}
\left\vert \sum_{n,m}\gamma _{nm}\mathrm{tr}[\rho (W_{1}^{(a_{n})}\otimes
W_{2}^{(b_{m})})]\right\vert \leq 2\left\Vert T_{112}\right\Vert _{1},
\label{36}
\end{equation}%
whenever $\gamma _{11}\gamma _{21}=-\gamma _{12}\gamma _{22}$. Here, $%
T_{122} $ and $T_{112}$ are any source-operators for a state $\rho .$
\end{proposition}

\begin{proof}
Due to the upper bounds (\ref{20}) and (\ref{21}), we have: 
\begin{eqnarray}
&&\left\vert \sum_{n,m}\gamma _{nm}\mathrm{tr}[\rho (W_{1}^{(a_{n})}\otimes
W_{2}^{(b_{m})})]\right\vert  \label{37} \\
&\leq &||T_{122}||_{1}\{2\text{ }+(\gamma _{11}\gamma _{12}+\gamma
_{21}\gamma _{22})\mathrm{tr}[\sigma _{T_{122}}(W_{2}^{(b_{1})}\otimes
W_{2}^{(b_{2})})]\text{ }\},  \notag \\
&&  \notag \\
&&\left\vert \sum_{n,m}\gamma _{nm}\mathrm{tr}[\rho (W_{1}^{(a_{n})}\otimes
W_{2}^{(b_{m})})]\right\vert  \label{38} \\
&\leq &||T_{112}||_{1}\{2+(\gamma _{11}\gamma _{21}+\gamma _{12}\gamma _{22})%
\mathrm{tr}[\sigma _{T_{112}}(W_{1}^{(a_{1})}\otimes W_{1}^{(a_{2})})]\text{ 
}\},  \notag
\end{eqnarray}%
and these relations prove the statement.
\end{proof}

\section{Validity of classical Bell-type inequalities in the quantum case}

Propositions 3 and 4 clearly indicate the cases where a bipartite quantum
state satisfies a classical CHSH-form inequality and the original Bell
inequality for \emph{any} bounded quantum observables. Notice that, in our
setting, bounded quantum observables may be of any spectral types\footnote{%
In a particular case, where either $\mathcal{H}$ is finite dimensional or,
in (\ref{39}), bounded quantum observables on infinite dimensional $\mathcal{%
H}$ have discrete spectra, the validity of a CHSH-form inequality for a 
\textit{DSO} state can be extracted from theorems 1 plus 2 in [6] - the
proof of theorem 2 in [6] is essentially built up on discreteness.}.

\begin{theorem}[DSO states and the CHSH inequality]
A density source-operator\footnote{%
See definition 2, section 2.1.} (DSO) state $\rho $ on $\mathcal{H}%
_{1}\otimes \mathcal{H}_{2}$ satisfies the original CHSH inequality [2]: 
\begin{equation}
\left\vert \mathrm{tr}[\rho (W_{1}^{(a_{1})}\otimes
W_{2}^{(b_{1})}+W_{1}^{(a_{1})}\otimes
W_{2}^{(b_{2})}+W_{1}^{(a_{2})}\otimes
W_{2}^{(b_{1})}-W_{1}^{(a_{2})}\otimes W_{2}^{(b_{2})})]\right\vert \leq 2,
\label{39}
\end{equation}%
for any bounded quantum observables $W_{1}^{(a_{n})},W_{2}^{(b_{m})},$ $%
n,m=1,2,$ with operator norms $||\cdot ||\leq 1$.
\end{theorem}

If a DSO state on $\mathcal{H}\otimes \mathcal{H}$ is symmetric then, for
this state, density source-operators $R_{\blacktriangleright }$ and $%
R_{\blacktriangleleft }$ exist simultaneously, and from proposition 4 there
follows:

\begin{theorem}
Let $\gamma _{nm},$ $n,m=1,2,$ be any real coefficients with $|\gamma
_{nm}|\leq 1$ such that $\gamma _{11}\gamma _{12}=-\gamma _{21}\gamma _{22}$
\ or $\ \gamma _{11}\gamma _{21}=-\gamma _{12}\gamma _{22}.$\newline
A symmetric DSO state $\rho $ on $\mathcal{H}\otimes \mathcal{H}$ satisfies
the extended CHSH inequality [3]: 
\begin{equation}
\text{\ }\left\vert \sum_{n,m=1,2}\gamma _{nm}\mathrm{tr}[\rho
(W_{1}^{(a_{n})}\otimes W_{2}^{(b_{m})})]\right\vert \leq 2,  \label{40}
\end{equation}%
for any bounded quantum observables $W_{1}^{(a_{n})},$ $W_{2}^{(b_{m})},$ $%
n,m=1,2,$ with operator norms $||\cdot ||\leq 1.$
\end{theorem}

Due to proposition 3, we have the following general statements on Bell class
states (cf. definition 3, section 2.1).

\begin{theorem}[Bell class states and the Bell inequality]
A Bell class state $\rho $ on $\mathcal{H}\otimes \mathcal{H}$ satisfies the
perfect correlation form of the original Bell inequality [1]: 
\begin{eqnarray}
&&\left\vert \mathrm{tr}[\rho (W_{1}\otimes W_{2})]-\mathrm{tr}[\rho
(W_{1}\otimes \widetilde{W}_{2})]\right\vert  \label{41} \\
&\leq &\mathrm{1}-\mathrm{tr}[\rho (W_{2}\otimes \widetilde{W}_{2})],  \notag
\\
&&  \notag \\
&&\left\vert \mathrm{tr}[\rho (W_{1}\otimes W_{2})]-\mathrm{tr}[\rho (%
\widetilde{W}_{1}\otimes W_{2})]\right\vert  \notag \\
&\leq &\mathrm{1}-\mathrm{tr}[\rho (W_{1}\otimes \widetilde{W}_{1})],  \notag
\end{eqnarray}%
for any bounded quantum observables with operator norms $||\cdot ||\leq 1.$
\end{theorem}

\begin{corollary}
Any Bell class state $\rho $ on $\mathcal{H}\otimes \mathcal{H}$ satisfies
the extended CHSH inequality (\ref{40}).
\end{corollary}

In the right-hand side of (\ref{41}), the operators can be interchanged.

It is necessary to underline that, in the physical literature, the validity
of the perfect correlation form of the original Bell inequality for a
bipartite state on $\mathcal{H}\otimes \mathcal{H}$ has been always linked
with Bell's assumption of perfect correlations if the same quantum
observable is measured on both sides (cf. in [1]).

In [3], we proved that separable states of the special form\footnote{%
As we discussed in section 2.1, these separable states belong to the Bell
class.} (49) in [3] satisfy (\ref{41}) for any bounded quantum observables
and do not necessarily exhibit perfect correlations. Theorem 3 generalizes
this our result in [3] and indicates that there exists the whole class of
bipartite states, separable and nonseparable, where each state satisfies the
perfect correlation form of the original Bell inequality\ for any three
bounded quantum observables and does not necessarily exhibit perfect
correlations.

In case of, for example, a dichotomic observable $W_{2}$, with eigenvalues $%
\pm 1,$ the latter means that a Bell class state $\rho $ satisfies (\ref{41}%
) even if the correlation function $\mathrm{tr}[\rho (W_{2}\otimes
W_{2})]\neq 1.$

Due to theorem 3 and proposition 2, \emph{the nonseparable Werner state }(%
\ref{9})\emph{\ on }$\mathbb{C}^{d}\otimes \mathbb{C}^{d},$ $\forall d\geq
3, $ \emph{satisfies the perfect correlation form of the original Bell
inequality for any bounded quantum observables and does not necessarily
exhibit perfect correlations.}

The upper bounds in proposition 3 allow us to introduce also a condition
sufficient for the validity of the original Bell inequality for a bipartite
state and some three quantum observables.

\begin{theorem}[General sufficient condition ]
If, for a DSO state $\rho $ on $\mathcal{H}\otimes \mathcal{H}$, there
exists a density source-operator $R_{\blacktriangleright }$ such that: 
\begin{equation}
\mathrm{tr}[\sigma _{R_{\blacktriangleright }}(W_{2}\otimes \widetilde{W}%
_{2})]=\pm \mathrm{tr}[\rho (W_{2}\otimes \widetilde{W}_{2})],\text{ \ \ }%
\sigma _{R_{\blacktriangleright }}=\mathrm{tr}_{\mathcal{H}%
}^{(1)}[R_{\blacktriangleright }],  \label{42}
\end{equation}%
for bounded quantum observables $W_{2}$ and $\widetilde{W}_{2}$ with
operator norms $||\cdot ||\leq 1,$ then this DSO state $\rho $ and these
quantum observables $W_{2}$, $\widetilde{W}_{2}$ satisfy the original Bell
inequality [1]: 
\begin{eqnarray}
&&\left\vert \mathrm{tr}[\rho (W_{1}\otimes W_{2})]-\mathrm{tr}[\rho
(W_{1}\otimes \widetilde{W}_{2})]\right\vert  \label{43} \\
&\leq &1\mp \mathrm{tr}[\rho (W_{2}\otimes \widetilde{W}_{2})],  \notag
\end{eqnarray}%
in its perfect correlation (minus sign) or anticorrelation (plus sign)
forms. Here, $W_{1}$ is any bounded quantum observable with $||\cdot ||\leq
1 $.
\end{theorem}

Notice that, in theorem 3, the sufficient condition concerns only a
bipartite state property and refers only to the perfect correlation form of
the Bell inequality. A Bell class state satisfies the (plus sign) condition (%
\ref{42}) for any observables $W_{2},$ $\widetilde{W}_{2}$.

In theorem 4, the sufficient condition (\ref{42}) establishes the
restriction on the combination - \textit{quantum observables and a DSO state}%
, and concerns both forms of the original Bell inequality. In general, a DSO
state satisfying the condition (\ref{42}) does not necessarily either belong
to the Bell class or satisfy (\ref{42}) for any $W_{2},$ $\widetilde{W}_{2}.$

For a symmetric DSO state, let us now prove that the sufficient condition (%
\ref{42}) is more general than Bell's perfect correlation/anticorrelation
restriction (\ref{44}) and includes the Bell restriction only as a
particular case.

\begin{proposition}
If a symmetric DSO state $\rho $ on $\mathcal{H}\otimes \mathcal{H}$
satisfies the Bell perfect correlation/anticorrelation restriction 
\begin{equation}
\mathrm{tr}[\rho (W_{2}\otimes W_{2})=\pm 1,  \label{44}
\end{equation}%
then this DSO state satisfies the sufficient condition (\ref{42}). The
converse is not true.
\end{proposition}

\begin{proof}
If a DSO state $\rho $ on $\mathcal{H}\otimes \mathcal{H}$ is symmetric then
it has both density source-operators, $R_{\blacktriangleleft }$ and $%
R_{\blacktriangleright }$. We have: 
\begin{eqnarray}
\mathrm{tr}[\sigma _{R_{\blacktriangleright }}(W_{2}\otimes \widetilde{W}%
_{2})] &=&\mathrm{tr}[R_{\blacktriangleright }(I_{\mathcal{H}}\otimes
W_{2}\otimes \widetilde{W}_{2})],  \label{45} \\
\mathrm{tr}[\rho (W_{2}\otimes \widetilde{W}_{2})] &=&\mathrm{tr}%
[R_{\blacktriangleright }(W_{2}\otimes I_{\mathcal{H}}\otimes \widetilde{W}%
_{2})],  \notag \\
\mathrm{tr}[\rho (W_{2}\otimes W_{2})] &=&\mathrm{tr}[R_{\blacktriangleright
}(W_{2}\otimes W_{2}\otimes I_{\mathcal{H}})].  \notag
\end{eqnarray}%
Using the arguments based on the von Neumann theorem [8] and quite similar
to ones in proposition 3, we derive: 
\begin{eqnarray}
\mathrm{tr}[\sigma _{T_{\blacktriangleright }}(W_{2}\otimes \widetilde{W}%
_{2})] &=&\int_{\mathbb{R}}\varphi _{2}(\xi )\varphi _{3}(\xi )\nu (d\xi
;R_{\blacktriangleright }),  \label{46} \\
\mathrm{tr}[\rho (W_{2}\otimes \widetilde{W}_{2})] &=&\int_{\mathbb{R}%
}\varphi _{1}(\xi )\varphi _{3}(\xi )\nu (d\xi ;R_{\blacktriangleright }), 
\notag \\
\mathrm{tr}[\rho (W_{2}\otimes W_{2})] &=&\int_{\mathbb{R}}\varphi _{1}(\xi
)\varphi _{2}(\xi )\nu (d\xi ;R_{\blacktriangleright }),  \notag
\end{eqnarray}%
where:\newline
(i) $\nu (\cdot ;R_{\blacktriangleright }):=\mathrm{tr}[R_{%
\blacktriangleright }P_{V}(\cdot )]$ is a probability distribution on $(%
\mathbb{R},$ $\mathcal{B}_{\mathbb{R}}),$ induced\footnote{%
See proposition 3.} by the projection-valued measure $P_{V}$ of a quantum
observable $V$ on $\mathcal{H}\otimes \mathcal{H}\otimes \mathcal{H}$
(corresponding, due to the von Neumann theorem, to three mutually commuting
observables $W_{2}\otimes I_{\mathcal{H}}\otimes I_{\mathcal{H}},$ $I_{%
\mathcal{H}}\otimes W_{2}\otimes I_{\mathcal{H}}$, $I_{\mathcal{H}}\otimes
I_{\mathcal{H}}\otimes \widetilde{W}_{2})$; \newline
(ii) $\varphi _{1}$, $\varphi _{2},$ $\varphi _{3}$ are bounded Borel
real-valued functions on $(\mathbb{R},$ $\mathcal{B}_{\mathbb{R}}),$ with
supremum norms $||\varphi _{n}||\leq 1,$ such that $\varphi
_{1}(V)=W_{2}\otimes I_{\mathcal{H}}\otimes I_{\mathcal{H}},$ $\varphi
_{2}(V)=I_{\mathcal{H}}\otimes W_{2}\otimes I_{\mathcal{H}}$ and $\varphi
_{3}(V)=I_{\mathcal{H}}\otimes I_{\mathcal{H}}\otimes \widetilde{W}_{2}.$ 
\newline
If $\rho $ satisfies the Bell restriction (\ref{44}) then, due to (\ref{46}%
): 
\begin{equation}
\int_{\mathbb{R}}\varphi _{1}(\xi )\varphi _{2}(\xi )\nu (d\xi
;T_{\blacktriangleright })=\pm 1.  \label{47}
\end{equation}%
The latter implies $\varphi _{1}(\xi )\varphi _{2}(\xi )=\pm 1$, $\nu $-$%
a.e. $ Since $||\varphi _{1}||,||\varphi _{2}||$ $\leq 1$, we have $\varphi
_{1}(\xi )=\pm \varphi _{2}(\xi )$, $\nu $-$a.e$, and, hence, 
\begin{eqnarray}
&&\mathrm{tr}[\sigma _{T_{\blacktriangleright }}(W_{2}\otimes \widetilde{W}%
_{2}\mp W_{2}\otimes \widetilde{W}_{2})]  \label{48} \\
&=&\int_{\mathbb{R}}\{\varphi _{2}(\xi )\mp \varphi _{1}(\xi )\}\varphi
_{3}(\xi )\nu (d\xi ;T_{\blacktriangleright })=0.  \notag
\end{eqnarray}%
The converse statement is not true and a DSO state, satisfying (\ref{48}),
does not necessarily satisfy (\ref{47}).
\end{proof}

\medskip

\emph{Thus, a DSO\ state satisfying the general sufficient condition }(\ref%
{42})\emph{\ does not necessarily exhibit Bell's perfect
correlations/anticorrelations.}

\subsection{Generalized quantum measurements of Alice and Bob}

In the physical literature, joint measurements on a bipartite system are
usually referred to as measurements of Alice and Bob. Theorems 1-4 and
proposition 5 specify the relations between the product expectation values
under projective quantum measurements of Alice and Bob.

To analyze the situation under \emph{generalized} joint quantum measurements
on a bipartite quantum state, let us recall that an Alice/Bob joint
generalized quantum measurement, with real-valued outcomes $\lambda _{1}\in $
$\Lambda _{1}$, $\lambda _{2}\in \Lambda _{2}$ of any type, is described by
the positive operator-valued (\emph{POV}) measure 
\begin{equation}
M^{(a,b)}(B_{1}\times B_{2})=M_{1}^{(a)}(B_{1})\otimes M_{2}^{(b)}(B_{2}),%
\text{ \ \ }\forall B_{1}\subseteq \Lambda _{1},\text{ }\forall
B_{2}\subseteq \Lambda _{2},  \label{49}
\end{equation}%
where $"a"$ and $\Lambda _{1}$ refer to a setting and an outcome set on the
side of Alice while \textquotedblright $b"$ and $\Lambda _{2}$ - on the side
of Bob. For simplicity, we further suppose $\left\vert \lambda
_{1}\right\vert \leq 1,$ $\left\vert \lambda _{2}\right\vert \leq 1$.

For a quantum state $\rho $ on $\mathcal{H}_{1}\otimes \mathcal{H}_{2},$ the
formula\footnote{%
See also [3], section 3.1.} 
\begin{align}
\langle \lambda _{1}\lambda _{2}\rangle _{\rho }^{(a,b)}& :=\int_{\Lambda
_{1}\times \Lambda _{2}}\lambda _{1}\lambda _{2}\mathrm{tr}[\rho
(M_{1}^{(a)}(d\lambda _{1})\otimes M_{2}^{(b)}(d\lambda _{2}))]  \label{50}
\\
& =\mathrm{tr}[\rho (W_{1}^{(a)}\otimes W_{2}^{(b)})]  \notag
\end{align}%
represents the expectation value of the product $\lambda _{1}\lambda _{2}$
of outcomes observed by Alice and Bob. Here, 
\begin{equation}
W_{1}^{(a)}:=\int_{\Lambda _{1}}\lambda _{1}M_{1}^{(a)}(d\lambda _{1}),\text{
\ \ }W_{2}^{(b)}:=\int_{\Lambda _{2}}\lambda _{2}M_{2}^{(b)}(d\lambda _{2})
\label{51}
\end{equation}%
are bounded quantum observables, with operator norms $||W_{1}^{(a)}||$ $\leq
1,$ $||W_{2}^{(b)}||\leq 1,$ representing, respectively, quantum averages on
the sides of Alice and Bob.

Theorems 1-3 and the representation (\ref{50}) imply:

\begin{theorem}
The product expectation values in a DSO state $\rho $ on $\mathcal{H}%
_{1}\otimes \mathcal{H}_{2}$ satisfy the original CHSH inequality: 
\begin{equation}
|\text{ }\langle \lambda _{1}\lambda _{2}\rangle _{\rho }^{(a_{1},b_{1})%
\text{ }}+\langle \lambda _{1}\lambda _{2}\rangle _{\rho }^{(a_{1},b_{2})%
\text{ }}+\langle \lambda _{1}\lambda _{2}\rangle _{\rho }^{(a_{2},b_{1})%
\text{ }}-\langle \lambda _{1}\lambda _{2}\rangle _{\rho }^{(a_{2},b_{2})%
\text{ }}|\text{ }\leq 2,  \label{52}
\end{equation}%
under any generalized quantum measurements (\ref{49}) of Alice and Bob with
outcomes $\left\vert \lambda _{1}\right\vert \leq 1,$ $\left\vert \lambda
_{2}\right\vert \leq 1$ of any type.
\end{theorem}

\smallskip

\begin{theorem}
Let $\gamma _{nm},$ $n,m=1,2,$ be any real coefficients with $\left\vert
\gamma _{nm}\right\vert \leq 1$ and $\gamma _{11}\gamma _{12}=-\gamma
_{21}\gamma _{22}$ \ or $\ \gamma _{11}\gamma _{21}=-\gamma _{12}\gamma
_{22} $. If a DSO state $\rho $ on $\mathcal{H}\otimes \mathcal{H}$ is
either symmetric or of the Bell class then the product expectation values in
this $\rho $ satisfy the extended CHSH inequality: 
\begin{equation}
\left\vert \sum_{n,m}\gamma _{nm}\langle \lambda _{1}\lambda _{2}\rangle
_{\rho }^{(a_{n},b_{m})}\right\vert \leq 2,  \label{53}
\end{equation}%
under any generalized quantum measurements (\ref{49}) of Alice and Bob with
outcomes $\left\vert \lambda _{1}\right\vert \leq 1,$ $\left\vert \lambda
_{2}\right\vert \leq 1$ of any type.
\end{theorem}

\smallskip

\begin{theorem}
Under any joint generalized quantum measurements (\ref{49}) of Alice and Bob
where: 
\begin{equation}
\int_{\Lambda _{1}}\lambda _{1}M_{1}^{(b_{1})}(d\lambda _{1})=\int_{\Lambda
_{2}}\lambda _{2}M_{2}^{(b_{1})}(d\lambda _{2}),  \label{54}
\end{equation}%
the product expectation values in every Bell class state $\rho $ on $%
\mathcal{H}\otimes \mathcal{H}$ satisfy the perfect correlation form of the
original Bell inequality: 
\begin{equation}
\left\vert \langle \lambda _{1}\lambda _{2}\rangle _{\rho
}^{(a,b_{1})}-\langle \lambda _{1}\lambda _{2}\rangle _{\rho
}^{(a,b_{2})}\right\vert \leq 1-\langle \lambda _{1}\lambda _{2}\rangle
_{\rho }^{(b_{1},b_{2})}.  \label{55}
\end{equation}
\end{theorem}

\medskip 

The operator relation (\ref{54})\ does not mean the perfect correlations of
outcomes on the sides of Alice and Bob and is always true in case of
projective Alice and Bob measurements of the same quantum observable on both
sides.

Theorem 4 can be also easily generalized to the case of joint generalized
quantum measurements.\bigskip 

\textbf{Acknowledgments}. I am grateful to Marek Bozejko, Klaus Molmer and
Asher Peres for valuable discussions.


\bigskip

\begin{thebibliography}{9}
\bibitem{1} J.S. Bell. \emph{Speakable and unspeakable in quantum mechanics.}
Cambridge University Press, Cambridge, England (1987)

\bibitem{2} J.F. Clauser, M.A. Horne, A. Shimony, and R.A. Holt. \emph{Phys.
Rev. Letters} \textbf{23}, 880-884 (1969)

\bibitem{3} E.R. Loubenets. \emph{Phys. Rev. A }\textbf{69}, 042102 (2004)

\bibitem{4} R.F. Werner. \emph{Phys. Rev. A }\textbf{40}, 4277\ (1989)

\bibitem{5} R.F. Werner and M.M. Wolf. \emph{Quantum information and
communication. }\textbf{1}, 1 (2001)

\bibitem{6} M. Terhal, A.C. Doherty, and D. Schwab (2003). \emph{Phys. Rev.
Lett.} \textbf{90}, 157903.

\bibitem{7} A. Peres. \emph{Phys. Rev. Lett}. \textbf{77, }1413 (1996)

\bibitem{8} J. Von Neumann. \emph{Mathematische Grundlagen der
Quantenmechanik}$.$ Springer-Verlag, Berlin (1932). English translation: 
\emph{Mathematical} \emph{Foundations of Quantum Mechanics}. Dover, New York
(1954)
\end{thebibliography}
\end{document}